\documentclass[aps, pra, letterpaper, onecolumn, superscriptaddress, notitlepage, 10pt]{revtex4-1}
\usepackage{times}

\usepackage{xcolor}
\usepackage{amsmath,amsfonts,amssymb}
\usepackage{graphicx}
\usepackage[caption=false]{subfig}
\usepackage{enumerate}
\usepackage{mathrsfs}
\usepackage{enumitem}

\usepackage{tikz}
\usetikzlibrary{decorations.pathreplacing,calc}

\usepackage[pdfpagelabels,pdftex,bookmarks,breaklinks]{hyperref}
\definecolor{darkblue}{RGB}{0,0,127} 
\definecolor{darkgreen}{RGB}{0,150,0}
\hypersetup{colorlinks, linkcolor=darkblue, citecolor=darkgreen, filecolor=red, urlcolor=blue}
\hypersetup{pdfauthor={Simon Burton, Courtney G. Brell, Steven T. Flammia}}
\hypersetup{pdftitle={Classical Simulation of Quantum Error Correction in a Fibonacci Anyon Code}}

\usepackage[normalem]{ulem}

\newcommand{\Fref}[1]{Fig.~\ref{#1}}

\newcommand{\e}{\mathrm{e}}
\newcommand{\vac}{\mathbb{I}}

\newcommand{\ket}[1]{|{#1}\rangle}

\newcommand{\bra}[1]{\langle{#1}|}
\newcommand{\ketbra}[2]{\ket{#1}\!\bra{#2}}

\newcommand{\proj}[1]{\ketbra{#1}{#1}}

\begin{document}

\title{Classical Simulation of Quantum Error Correction in a Fibonacci Anyon Code}

\author{Simon Burton}
\affiliation{Centre for Engineered Quantum Systems, School of Physics, 
The University of Sydney, Sydney, Australia}
\author{Courtney G.\ Brell}
\affiliation{Institut f\"{u}r Theoretische Physik, Leibniz Universit\"{a}t Hannover, 
Appelstra\ss{}e 2, 30167 Hannover, Germany}
\affiliation{Perimeter Institute for Theoretical Physics, 
Waterloo, Canada}
\author{Steven T.\ Flammia}
\affiliation{Centre for Engineered Quantum Systems, School of Physics, 
The University of Sydney, Sydney, Australia}

\date{\today}

\begin{abstract}
Classically simulating the dynamics of anyonic excitations in two-dimensional quantum systems is likely intractable in general because such dynamics are sufficient to implement 
universal quantum computation. However, processes of interest for the study of quantum 
error correction in anyon systems are typically drawn from a restricted class that displays 
significant structure over a wide range of system parameters.
We exploit this structure to classically simulate, and thereby demonstrate the success of, an 
error-correction protocol for a quantum memory based on the universal Fibonacci anyon 
model.  We numerically simulate a phenomenological model of the system and noise 
processes on lattice sizes of up to 
$128\times128$ sites, and find a lower bound on the error-correction threshold of 
approximately $0.125$ errors per edge, which is comparable to those previously known for abelian and 
(non-universal) nonabelian anyon models.
\end{abstract}

\maketitle


Topologically ordered quantum systems in two dimensions show great promise for 
long-term storage and processing of quantum information~\cite{Kitaev2003, Dennis2002, Nayak2008}. 
The topological features of such systems are insensitive to local 
perturbations~\cite{Bravyi2010, Bravyi2011a, Michalakis2013}, and they have quasiparticle excitations 
exhibiting anyonic statistics~\cite{Wilczek1990}. These systems can be used as 
quantum memories~\cite{Kitaev2003, Dennis2002} or to perform universal topological 
quantum computation~\cite{Freedman2002, Nayak2008}.

Quantum error correction is vital to harnessing the computational power of topologically 
ordered systems. When coupled to a heat bath at any non-zero temperature, thermal fluctuations 
will create spurious anyons that diffuse and quickly corrupt the stored quantum 
information~\cite{Pastawski2010}. Thus, the passive protection provided by the mass gap 
at low temperature must be augmented by an \emph{active} decoding procedure. 

In order to efficiently classically simulate an error-correction protocol for 
a topologically ordered quantum memory, it is necessary to simulate 
the physical noise processes, the decoding algorithm, and the physical recovery operations. 
Decoding algorithms are typically designed to run efficiently on a 
classical computer, but there is generally no guarantee that the 
noise and recovery processes should be classically simulable.
Because of this, almost all of the sizable research effort 
on active quantum error correction for topological systems has focused 
on the case of abelian anyons~\cite{Dennis2002, Duclos-Cianci2010, Duclos-Cianci2010a, Wang2010, Wang2010a, Duclos-Cianci2013, Bravyi2011, Bombin2012, Wootton2012, Anwar2014, Watson2014, Hutter2014a, Bravyi2014, Wootton2015, Fowler2015, Andrist2015}, which can be efficiently simulated due to the fact that they cannot be used for quantum computation.

Recent investigations have begun to explore quantum error correction for nonabelian anyon 
models~\cite{Brell2013, Wootton2013, Hutter2014, Wootton2015b, Hutter2015continuous}. Nonabelian anyon models are especially interesting 
because braiding and fusion of these anyons in general allows for the implementation of universal quantum 
computation. However, the initial studies of error-correction in nonabelian anyon systems have focused on specific models, such as the Ising 
anyons~\cite{Brell2013, Hutter2015continuous} and the so-called $\Phi$-$\Lambda$ 
model~\cite{Wootton2013, Hutter2014} that, while nonabelian, are not universal for quantum computation. The general dynamics of these particular anyon models is known to be efficiently classically simulable, a fact
that was exploited to enable efficient simulation of error correction 
in these systems. When considering more general anyon models, their 
ability to perform universal quantum computation would seem a significant 
barrier to their simulation on a classical computer. While simulation 
of general dynamics does indeed seem intractable, we argue that 
the kinds of processes that are typical of thermal noise 
are sufficiently structured  to allow for their classical simulation 
in the regimes where we expect successful error correction to 
be possible. This insight allows us to simulate the noise 
and recovery processes for a quantum code based on a universal anyon model.

Concretely, we consider quantum error correction in a two-dimensional 
system with Fibonacci anyons, a class of nonabelian anyons that are universal for quantum 
computation~\cite{Freedman2002, Nayak2008}. Fibonacci anyons are experimentally motivated as the 
expected excitations of the $\nu=\frac{12}{5}$ fractional quantum Hall 
states~\cite{Slingerland2001}, and can be realized in several spin 
models~\cite{Levin2005, Bonesteel2012, Kapit2013, Palumbo2014} and composite 
heterostructures~\cite{Mong2014}.
Any of these physical systems could be used to perform universal topological quantum computation, and can be modelled by our simulations. Natural sources of noise from thermal fluctuations or external perturbations will be suppressed by the energy gap but must still be corrected to allow for scalable computation.

We use a flexible phenomenological model of dynamics and thermal 
noise to describe a system with Fibonacci anyon excitations. Within 
this model, we apply existing general topological error-correction protocols, and 
simulate the successful preservation of quantum information encoded in topological 
degrees of freedom. Topological quantum computation protocols using nonabelian anyons 
typically implicitly assume the existence of an error-correction protocol to 
correct for diffusion or unwanted creation of anyons. 
The ability to simulate the details of how and when these techniques succeed on finite system sizes has not previously been available, and so our results are the first explicit demonstration that such a scheme will be successful when applied to a universal topological quantum computer.

We will first introduce our setting and notation by giving a brief review of the background of Fibonacci anyons and error-correction for such systems in Section \ref{s:background}, before sketching an argument for why we expect these dynamics to be classically simulable in the regime of interest and outlining our simulation algorithm in Section \ref{s:simulation}. We then present numerical results and a discussion in Section \ref{s:results}. A more detailed explanation of our model, simulation algorithm, and error-correction protocol follows in Section \ref{s:technical}.


\section{Background}\label{s:background}

\subsection{Topological model}

\begin{figure}[t!]
\begin{center}
    \includegraphics[]{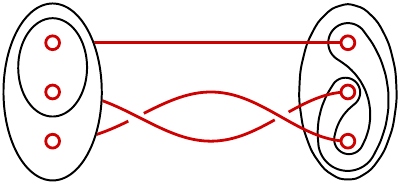}
\caption{
The deformation of observables corresponding to the implementation of a braid process on three anyons is shown. Configurations before and after are connected by world lines shown in red, while the charge observables are shown in black.
This process relates two different ways of fusing the top two charges.
\label{f:braidloop}
}
\end{center}
\vspace{-10pt}
\end{figure}

We consider a 
2-dimensional system with
point-like anyonic quasiparticle excitations.
The worldlines of these particles form \emph{braids}
in (2+1)-dimensions
which act unitarily on the Hilbert space of degrees of freedom known as the \emph{fusion space}.
This space is associated with
the topological observables of the system: these
are total charge measurements within regions bounded by 
closed loops.
The possible results of such a charge measurement correspond to different anyon particle types in the model.
These dynamics and observables obey algebraic rules
given by a unitary modular tensor category~\cite{Wang2010b}. In contrast to the approach often taken, we consider the topological observables, rather than the anyons themselves, to be affected by braiding processes (see \Fref{f:braidloop}).

The defining difference between abelian and nonabelian
anyon theories is that in an abelian theory particle content alone uniquely determines the outcome of joint charge measurements.
In contrast, outcomes for nonabelian charge measurements depend on the history of the particles as well as their type.
That is, the fusion space of a set of abelian anyons is one-dimensional, while it is generally larger for nonabelian anyons.

We consider a system supporting nonabelian Fibonacci
anyon excitations, denoted by $\tau$.
Two such anyons can have total charge
that is either $\tau$ or $\vac$ (vacuum), or any superposition of these, and
so the fusion space in this case is 2-dimensional.
We can represent basis states for this space using diagrams 
of definite total charge for the Wilson loops, 
and arbitrary states as linear combinations of these diagrams:
\begin{align*}
\includegraphics[]{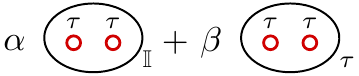}
\end{align*}
For $n$ anyons of type $\tau$, the dimension of the fusion space grows asymptotically as $\varphi^n$, where $\varphi=\frac{1+\sqrt{5}}{2}$ is the golden ratio.

Observables associated to non-intersecting loops commute, and so
a basis for the space can be built from a maximal
set of disjoint, nested loops.
For three anyons, two
possible ways of nesting loops are related
via $F$-moves:
\begin{align}
\label{e:fmove}
\includegraphics[]{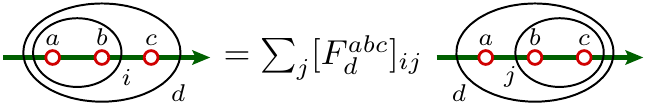}
\end{align}
Here we have shown particles with charges $a$, $b$, and $c$
as well as total charge $d$.
The lefthand side shows a state where the anyons with charge $a$ and $b$ are
observed to have total charge $i$.
The righthand side involves a superposition 
over total charge $j$ of the anyons with charge $b$ and $c$. 
The green line serves to linearly
order the anyons, and track
the history of braiding processes (see Ref.~\cite{Burton2016} for a more comprehensive discussion). It also denotes the direction along with $F$-moves occur.
A clockwise braiding process, or \emph{half-twist}, is effected by an $R$-move:
\begin{align}
\label{e:rmove}
\includegraphics[]{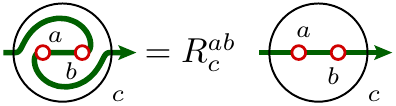}
\end{align}

These diagrams can be composed (glued)
by regarding the outermost observable
as a charge in another diagram.
This operation distributes over superpositions, and it also respects the $R$- and $F$-moves.
Conversely, \emph{fusion} of anyons corresponds to 
replacing the interior of a loop by a single charge.

For Fibonacci anyons, the non-trivial $R$ and $F$ moves are 
\begin{equation*}
	R_{\vac}^{\tau\tau} = \e^{\frac{-4\pi i}{5}} 
	\ \ , \ \
	R_\tau^{\tau\tau}= \e^{\frac{3\pi i}{5}} 
	\ \ , \ \
	F_{\tau}^{\tau\tau\tau} = \begin{pmatrix}\varphi^{-1}&\varphi^{-\frac{1}{2}}\\\varphi^{-\frac{1}{2}}&-\varphi^{-1}\end{pmatrix} \,,
\end{equation*}
where the matrix is given in a basis labelled $(\vac,\tau)$. For more details of the Fibonacci anyon theory, see e.g.\ Ref.~\cite{Nayak2008} and references therein. 

In addition to the fusion space there is also a global degeneracy associated with the topology of the 
manifold on which the anyons reside. We consider systems with the topology of a torus, 
which for the Fibonacci anyons gives rise to a 2-fold degeneracy.
This extra degree of freedom can be thought of being associated with the total anyonic charge (or flux) running through the torus itself. Since there are two different possible charges in the Fibonacci anyon model, the global degeneracy is 2.

\subsection{Noise and error-correction}

We consider encoding a qubit of quantum information in the global degeneracy associated 
with the topology of the manifold of our system.
We endow this manifold with a finite set of $L\times L$ Wilson loops arranged in
a square lattice tiling $\Lambda$.
These \emph{tiles} will be the observables accessible to the error-correction procedure (\emph{decoder}).
We can use these observables to construct an idealized Hamiltonian for this system of the form
 $H=-\sum_{t\in \Lambda}\proj{\vac}_t\;,$ 
with $\proj{\vac}_t$ the projector to charge $\vac$ at tile $t$, i.e.~the ground 
space of the model has vacuum total charge in each tile.
Given the toroidal boundary 
conditions, this space is the two-fold degenerate codespace.

Typical thermal noise processes in this kind of system are 
pair-creation, hopping, exchange etc.~of anyons. Such a model was analyzed in detail for Ising anyons in Ref.~\cite{Brell2013}.
For convenience, in the numerical results presented in this study we will restrict to considering pair-creation events only, though all of our simulation techniques could be applied directly to the more general noise considered in Ref.~\cite{Brell2013} as desired. It was seen in Ref.~\cite{Brell2013} that the pair-creation-only setting was sufficient to capture the qualitative features of an error-correction simulation for the Ising anyons and we have no reason to expect that this would change when considering the Fibonacci anyon model. 

Pair-creation events acting within a single tile do not affect the corresponding observable, and so the only pair-creation events we need explicitly consider act between neighboring pairs of tiles. These are then each associated with an edge of the lattice $\Lambda$, chosen 
uniformly at random to model high-temperature, short timescale thermalizing dynamics.
In studies of codes based on abelian anyon models, the error correction threshold or memory lifetime is typically quoted in terms of an iid noise strength per edge. This measure is not appropriate in a nonabelian anyon setting, where noise processes on neighbouring edges do not commute. We thus measure the error correction threshold and memory lifetime in terms of the average number of noise processes per edge. 
It is easy to see that, in those cases where they are both appropriate, these two methods coincide in the limit of low error rates, and for the typical noise strengths we encounter in this paper, the discrepancy is around $10\%$ (see Ref.~\cite{Brell2013} for more details). 

We consider this treatment of anyon dynamics to be a phenomenological model in that
it neglects any microscopic details of the system.
This is consistent with the principles of topologically 
ordered systems and anyonic physics, where the key universal features describing the 
anyon model correspond to large length-scale physics, while the microscopic physics plays 
a less important (and non-universal) role. Note also that our analysis applies equally well to either a continuum setting, or to discrete lattice models supporting anyonic excitations.

In order to perform a logical error on our code, a noise process must have support on a 
a homologically nontrivial region of the manifold.
These correspond to processes in which anyonic charge is transported around a non-trivial loop before 
annihilating to vacuum.

\begin{figure}[t!]
\begin{center}
	\includegraphics[width=0.7\columnwidth]{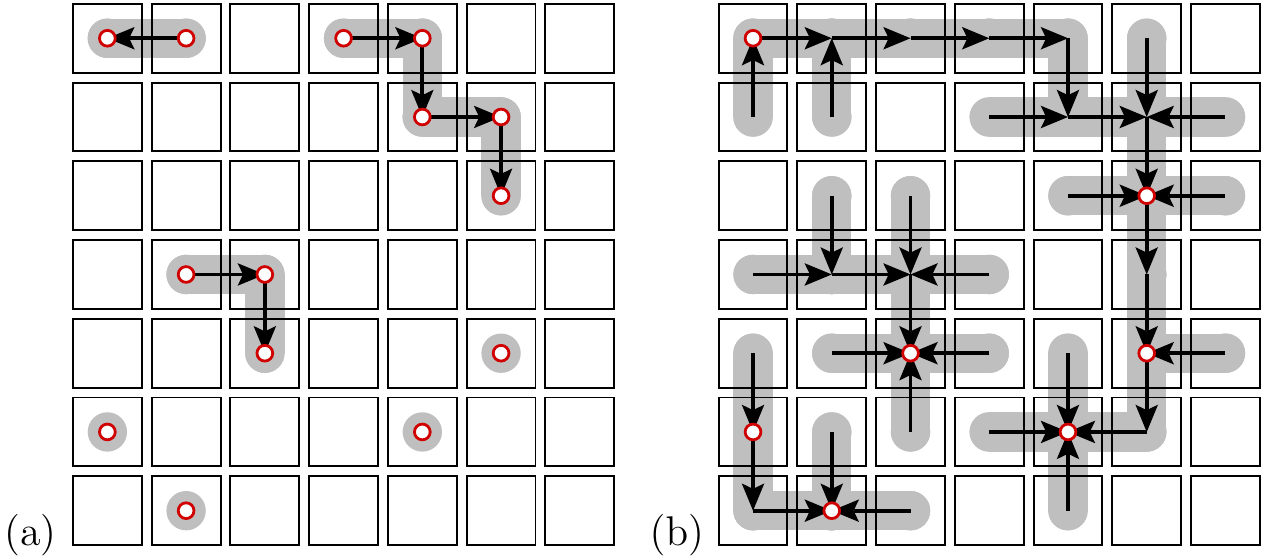}
\caption{The decoder works by maintaining a set of disjoint clusters as rooted trees.
(a) At the initial clustering stage, these trees are formed 
from neighboring sites that contain charges. Within each cluster, the 
charges are transported to the root of the tree (chosen 
arbitrarily), and their combined charge measured. The direction of transport 
(towards the root) is denoted by arrows.
(b) At each successive round, all trees are grown in 
every direction, and overlapping trees are joined. Again, any charges 
within a cluster are transported to the root of the 
tree and measured. All clusters with vacuum total charge are deleted.
\label{f:decode}
}
\end{center}
\vspace{-10pt}
\end{figure}

Our error-correction algorithm begins by measuring the charge on 
each tile of the lattice, producing a \emph{syndrome} 
of occupied sites. Following this, it joins nearby occupied tiles into clusters, and measures the total charge within each cluster.
Clusters with trivial charge are discarded, and then nearby clusters are merged (agglomeratively~\cite{Hastie2009}).
The merging process iterates at linearly increasing length scales, at each step measuring total charge and discarding trivial clusters, before concluding when 
there is at most a single cluster remaining (see \Fref{f:decode}).
This decoder is based on a hierarchical clustering algorithm~\cite{Hastie2009, Wootton2015b},
and follows a similar strategy to the hard-decision renormalization group decoder~\cite{Bravyi2011}. 
The error-correction protocol and simulation procedure are discussed in more depth in Section \ref{s:technical}.

Note that unlike the case of abelian anyons, the decoder
cannot determine the charge of each cluster given only the syndrome information (as in \cite{Bravyi2011}),
and so must repeatedly physically query the system to measure these charges.
Our simulation proceeds in this way as a dialogue between decoder
and system, terminating when either the decoder itself terminates or 
when it performs a homologically nontrivial operation, thereby performing a potential logical error.
Note that, as was found in Ref.~\cite{Brell2013}, we expect that our qualitative results may be reproduced by most alternative families of decoders, although it may not be clear in all cases how to extend these decoders to the nonabelian setting. The advantage of using a clustering decoder is its simplicity and flexibility, and the fact that its clustering scheme is compatible with the structure of the noise processes that allows us to classically simulate them.

\section{Simulation}\label{s:simulation}

\subsection{Classical simulability}

Although simulating pair creation, braiding, and fusion of Fibonacci anyons is equivalent 
in computational power to universal quantum computing (and thus unlikely to be classically 
tractable), noise processes and error-correction procedures have structure that we can 
exploit to efficiently simulate typical processes of interest. In particular, those 
processes in which we expect error-correction to succeed are also those that we expect to 
be able to efficiently simulate for the following heuristic reasons 
(we leave a more rigorous analysis of simulability for noise 
and error correction processes as an open problem).

Below the (bond) percolation threshold for (say) a 2D square lattice, we expect random sets of 
bonds to decompose into separate connected components 
of average size $O(\log(L))$ and variance $O(1)$~\cite{Bazant2000}.
Each noise process in our model is associated with a (randomly distributed) edge, and so 
disconnected components correspond to sets of anyons that could not have interacted at any 
point in their history. 
We are free to neglect the degrees of freedom associated with braiding between components 
because each component has trivial total charge.
This allows us to simulate the braiding processes within each component separately. 
In other words: the quantum state in the fusion space of all anyons factorizes into 
a tensor product over components. 
Since each 
component has size only $O(\log(L))$, we can typically simulate these dynamics efficiently 
because the resulting fusion space has dimension $O(\mathrm{poly}(L))$. 
Of course, this is only a probabilistic statement, and so there may be cases where there exist components with size larger than $O(\log(L))$. However, the likelihood of this is suppressed exponentially in the lattice size $L$~\cite{Grimmett1989}. 

However, random noise processes are not the only dynamics that we need to consider. We 
must also consider the effect of the error-correction routine itself. This acts iteratively to fuse 
anyons on increasing length scales. While this kind of fusion would typically merge components, 
forcing us to compute dynamics of larger and larger sets of anyons,
large components are sparsely distributed
(and thus unlikely to be merged), and in addition at each length scale the total number of 
anyons present is dramatically reduced by fusion, leading to a smaller number of anyons that 
must be simulated.

In the regime where the combined action of noise and error-correction does not percolate
it is reasonable to expect that the simulation is efficient.
However, with strong enough noise the state in the fusion space will
no longer decompose at all and computing dynamics will
become exponentially difficult in the system size.
Nevertheless, we are able to simulate error-correction in
the regime around the error-correction threshold for linear lattice sizes up to $L=128$.

\subsection{Simulation algorithm}

\begin{figure}[t!]
\begin{center}
	\includegraphics[width=0.7\columnwidth]{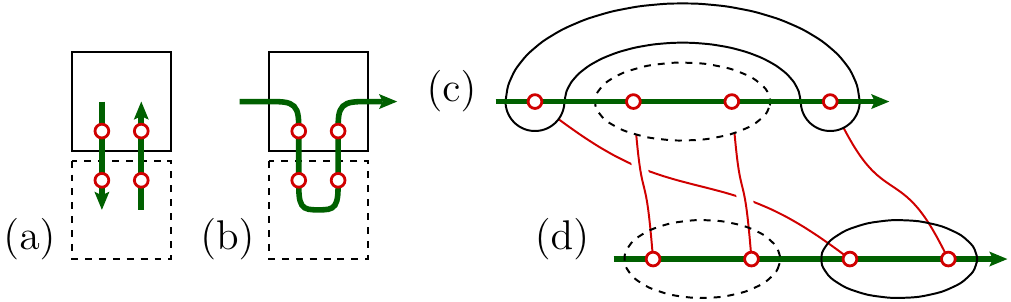}
\caption{
(a) Noise processes initially form isolated sets of pair-created anyons, 
each crossing the boundary of a tile. 
(b) To measure the total charge 
contained within each tile, 
we first join the participating curve 
diagrams arbitrarily into a single curve diagram.
(c) The total charge of each tile can then be found 
by braiding anyons around each other until all charges within 
a tile are neighbors on the curve, as in (d).  
The red lines correspond to the worldlines for these braids.
}
\label{f:syndrome}
\end{center}
\vspace{-10pt}
\end{figure}

In order to track the state of the system we represent the state of the fusion space with 
a set of disjoint directed curves, 
one for each connected component with trivial total charge.
Unlike previous work~\cite{Brell2013}, this creates a dynamically generated basis for the fusion space 
that allows for tensor factorization of disjoint components.
A pair-creation noise process corresponds to the 
addition of an extra curve to the lattice, supporting
two new anyons.
Following the noise processes, our simulation must 
measure the total charge within each tile; the results of 
these measurements will form the error syndrome. 
This requires joining the curves that intersect that tile, 
and then performing $R$- and $F$-moves on the resulting
curve so that 
all anyons within the tile have been 
localized within a contiguous region of the curve. 
An example of this 
procedure for a simple noise process is shown in \Fref{f:syndrome}.
This procedure is discussed in more detail in Section \ref{s:technical}, and the theoretical framework underlying it is expanded in Ref.~\cite{Burton2016}.
Braiding processes that encircle 
a non-trivial loop of our (toric) manifold can also be 
treated in an analogous way, following e.g.~Ref.~\cite{Pfeifer2012}.

Following the results of the charge measurements, the decoder determines 
a recovery operation that involves fusion of subsets of anyons. 
These fusions can be simulated and their results calculated in the same way, 
and the output charge placed at an appropriate point in the lattice. 
This procedure is iterated until either the decoder terminates successfully, 
or the simulation itself declares failure because the components become too large to analyze (roughly when they span the lattice).


\section{Results}\label{s:results}

\subsection{Numerical results}

\begin{figure}[t!]
\begin{center}
	\includegraphics[width=0.7\columnwidth]{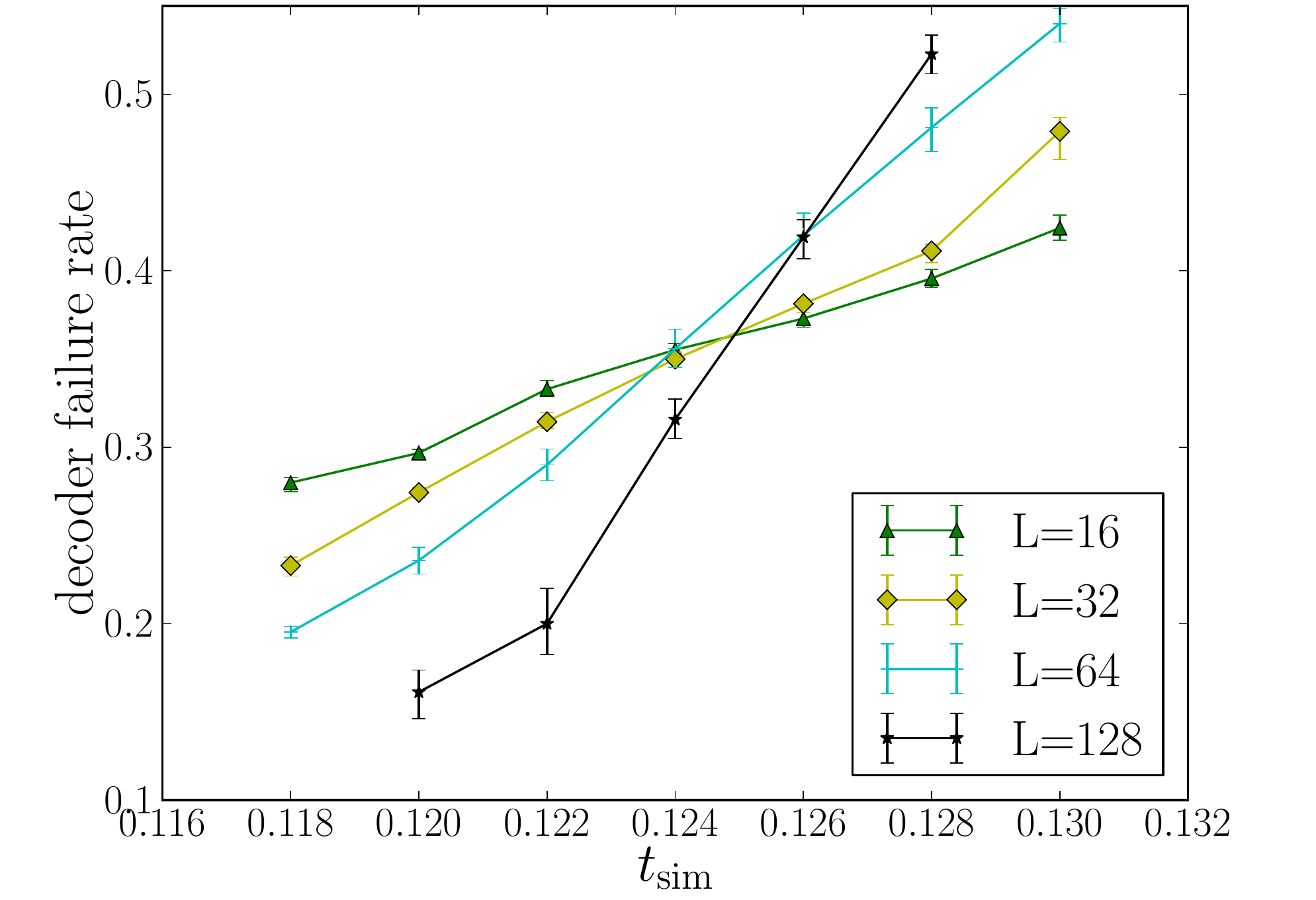}
\caption{The decoder failure rate (a lower bound on the logical error probability) is shown as a function of simulation time for linear lattice sizes from $L=16$ to $128$. 
This provides strong evidence of threshold behavior at a critical memory lifetime of $t_{\mathrm{sim}}^*\simeq 0.125$. 
The existence of such a threshold implies that the Fibonacci anyon code simulated here is able to perfectly reliably store quantum information for times less than $t_{\mathrm{sim}}^*$ in the $L\to \infty$ limit.}
\label{f:threshold}
\end{center}
\vspace{-10pt}
\end{figure}

We plot the performance of the decoder as a function of noise strength for varying lattice sizes in 
\Fref{f:threshold}. 
The noise strength is parameterized by the Poisson process duration $t_{\mathrm{sim}}$, representing the expected number of errors per edge during the simulation. 
We find evidence of a decoding threshold below which decoding succeeds with asymptotic 
certainty as the system size increases at $t_{\mathrm{sim}}\simeq 0.125 \pm 0.003$.

We can guarantee that error-correction will succeed whenever
the action of noise plus decoder 
does not percolate the anyons over the lattice,
but it is possible that percolated events may still result in no error. 
The connection between the percolation threshold and the error-correction threshold 
is not well understood in general~\cite{Hastings2014}, though it is clear 
that our threshold estimate would be a lower bound for 
the true threshold that may be found if all events 
(including those that have percolated) were simulated. 
In any case, we do not expect this possible discrepancy to be significant in our simulations.

\subsection{Discussion}

We have demonstrated classical simulation of successful error correction in a universal anyon model. 
Though we have chosen several properties of our model and 
simulation in a convenient way for simplicity,
Ref.~\cite{Brell2013} presents good evidence that it is
unlikely these choices will affect our results qualitatively.
In particular, although we have considered a logical qubit encoded in the global topological degrees of freedom of our 
system, we could have encoded it in the fusion space of several preferred anyons. 
This situation would be appropriate to model error-correction routines for topological quantum computation. 
Additionally, we expect our results to be stable to changes in details of the noise model and decoding algorithm, again following Ref.~\cite{Brell2013}.

None of our techniques are restricted to simulation of Fibonacci anyon dynamics, and could 
equally well be used to simulate successful error-correction protocols in an arbitrary anyon code. 
As such, our methods could be used to demonstrate successful 
error-correction for an arbitrary anyonic topological quantum computer.

There are several interesting avenues for further research. 
Although it is not completely obvious how to do so, 
applying similar methods to more realistic models such as concrete 
microscopic spin models or models with non-topological features would give 
direct insight into practical error rates needed for nonabelian topological quantum memories.
It would also be interesting to find \emph{classical} spin models 
whose phase diagram encodes the threshold for error correction in these systems~\cite{Dennis2002}.
Finally, an important extension of this work would be to the simulation of \emph{fault-tolerant} error-correction protocols for nonabelian anyon codes.


\section{Technical details}\label{s:technical}

\newcommand{\ntikzmark}[2]{#2\thinspace\tikz[overlay,remember picture,baseline=(#1.base)]{\node[inner sep=0pt] (#1) {};}}

\newcommand{\makebrace}[3]{%
    \begin{tikzpicture}[overlay, remember picture]
        \draw [decoration={brace,amplitude=0.5em},decorate]
        let \p1=(#1), \p2=(#2) in
        ({max(\x1,\x2)}, {\y1+0.8em}) -- node[right=0.6em] {#3} ({max(\x1,\x2)}, {\y2});
    \end{tikzpicture}
}

In order to estimate the error-correction threshold of Fibonacci anyon codes, we perform monte-carlo simulations of noise and recovery operations on systems of various sizes. For a given sample, the simulation can be broken down into several stages:
\begin{itemize}[noitemsep, topsep=0pt]
\item \ntikzmark{N}{Noise}
\item \ntikzmark{S}{Syndrome extraction}
\item \ntikzmark{D}{Decoding}
\item \ntikzmark{R}{Recovery}
\end{itemize}
\makebrace{S}{R}{Error-correction}
After application of noise, the subsequent three steps (forming the error-correction routine) are iterated until success or failure, as will be described in more detail below.

It should be noted that while the noise, syndrome extraction, and recovery steps of this procedure are intended to really be a simulation of the physics of an anyon system, the decoding step is better thought of as an implementation of the classical processing that a successful error-correction experiment will require. Furthermore, the simulation of the noise and recovery steps will be rendered straightforward by the way we model our system in the first place. As such, the main novelty of our simulations is in the model itself, and the syndrome extraction step.

The key to simulating these processes effectively is the ability to describe topological operations by \emph{curve diagrams}. This is an alternative representation of topological operations that is particularly suited to these kinds of numerical simulations of sparsely populated 2-dimensional anyon systems. The theoretical framework underpinning this description and the relation to to other representations of topological operations is discussed in Ref.~\cite{Burton2016}.

In this section, we first describe our physical model of an anyon system and the corresponding noise, before showing how to simulate joint charge measurements (the core of syndrome extraction) within this model. The use of curve-diagrams is presented in an intuitive but ad-hoc way. Finally we describe our decoder and detail holistically how the different parts of the simulation interact.

\subsection{Physical model}

We consider anyon systems defined on a torus.
We endow this with a $L\times L$ square lattice of observables:
$$
    \Lambda := \bigl\{ \gamma_{ij} \bigr\}_{i,j=1,...,L}
$$
These observables are the physically accessible charge observables that will be measured during syndrome extraction, before being passed to the decoder which will in turn determine physical recovery operations to be performed on the system.
We call each $\gamma_{ij}$ a \emph{tile.}
In figures we show a small gap between the tiles but this is not meant
to reflect an actual physical gap.

As noted above, the key to our model is to represent anyons and topological operations with curve diagrams. A curve diagram is a directed path that connects several anyons, which are denoted by preferred points along these paths. A curve diagram specifies a linear ordering of the anyons that allows us to define a basis for the corresponding fusion space, as well as to keep track of topological operations on this ordering. See Ref.~\cite{Burton2016} for a more formal introduction to curve diagrams.

The main noise processes we consider in our simulations are pair-creation events, though other topological processes such as hopping, exchange, etc.~can also be implemented straightforwardly.
The pair-creation process acts to populate the manifold with
a randomly distributed set of anyon pairs,
whose separation is much smaller than the size of a tile.

Each such pair will have vacuum total charge and so the observables
$\gamma_{ij}$ will only be affected by pairs that cross a tile boundary, i.e.~we
need only consider distributing these pairs along edges of the tiles:
\begin{center}
\includegraphics[width=0.3\columnwidth ]{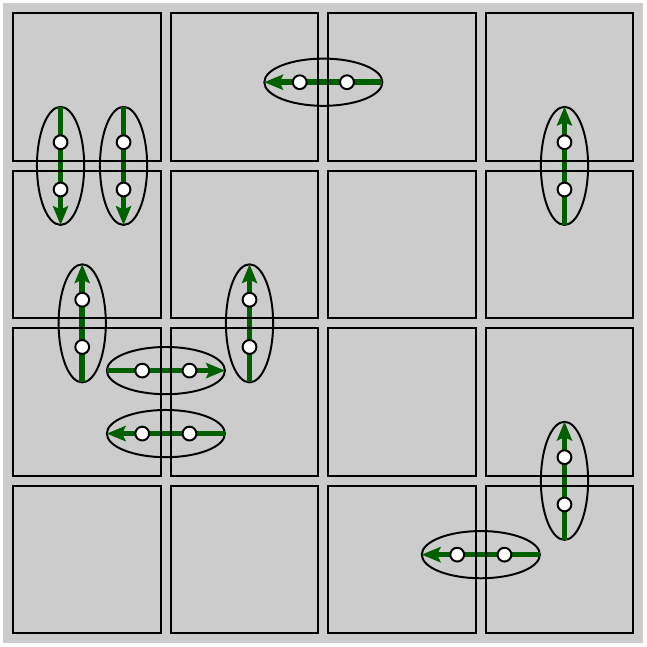}
\end{center}

In simulating system undergoing noise for time $T$, we place a Poisson-distributed number of pair-creation events on the edges of tiles, with mean $2L^2T$.

In order to allow us to later calculate fusion outcomes for sets of anyons, in addition to locations of created anyons we also assign each pair a (directed) piece of curve, denoted in green. This curve can be thought of as a choice of fusion basis for the anyons along it. Since the total charge of any newly-created particle-antiparticle pair is vacuum, each of their fusion spaces can be treated independently. This results in our being able to consider independent disconnected curves for each such pair. We only need consider the joint state in the fusion space of different sets of anyons if they have interacted, and so it is with curve diagrams. Using this picture, it is easy to track the effects of moving anyons, which essentially are represented by appropriate extensions or deformations of the curve along which the anyon lies. These kinds of straightforward processes are all that are required to describe the noise and recovery steps of our simulation.

\subsection{Syndrome extraction}

Syndrome extraction in this model corresponds to simulating measurements of the observables $\gamma_{ij}$, which represent the total charge contained within the corresponding tile.
In order to compute measurement outcomes for the $\gamma_{ij}$
we first need to concatenate any two curve diagrams that 
participate in the same $\gamma_{ij}.$
Because each curve has vacuum total charge this can be
done in an arbitrary way:
\begin{center}
\includegraphics[width=0.3\columnwidth ]{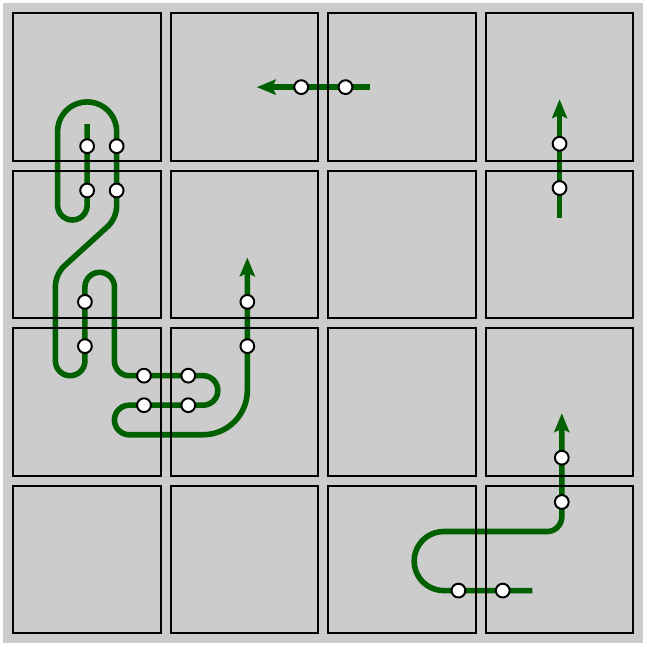}
\end{center}

Working in the basis picked out by the resulting curve
diagrams, we will show how to calculate measurement outcomes for each tile,
the result of which is recorded on the original curve:
\begin{center}
\includegraphics[width=0.3\columnwidth ]{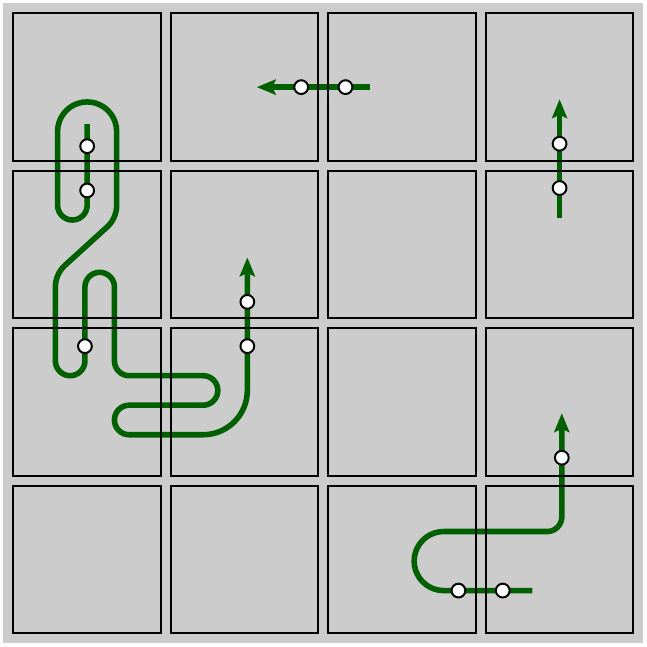}
\end{center}

%
%

The basic data structure involved in the
simulation we term a \emph{combinatorial curve diagram.}
Firstly, we will require each curve to intersect 
the edges of tiles transversally,
and in particular a curve will not touch a tile corner.

For each tile in the lattice,
we store a combinatorial
description of the curve(s) intersected with that tile.
Each component of such an intersection we call a \emph{piece-of-curve.}
\begin{center}
\includegraphics[]{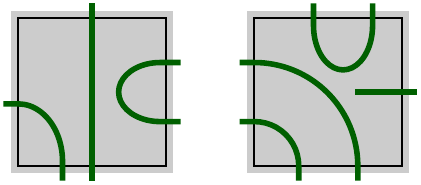}
\end{center}

We follow essentially the same approach as taken in \cite{Abramsky2007} 
to describe elements of a Temperley-Leib algebra, but
with some extra decorations.
The key idea is to store a \emph{word} for each tile, comprised of
the letters $\bigl<$ and $\bigr>$.
Reading in a clockwise direction around the edge of
the tile from the top-left corner,
we record our encounters with each piece-of-curve,
writing~$\bigl<$ for the first encounter, and~$\bigr>$ for the
second.
We may also encounter a dangling piece-of-curve
(the head or the tail), so we use another symbol $*$ for this.
The words for the above two tiles will then be 
$\bigl<\bigl<\bigr>\bigr>\bigl<\bigr>$ and $\bigl<\bigr>*\bigl<\bigl<\bigr>\bigr>.$
When the brackets are balanced,
each such word will correspond one-to-one with an intersection
of a curve in a tile, up to a continuous deformation of the interior of the tile,
i.e.~the data structure 
will be insensitive to any continuous deformation of the interior of the tile,
but the simulation does not need to track any of these degrees of freedom.

\begin{center}
\includegraphics[]{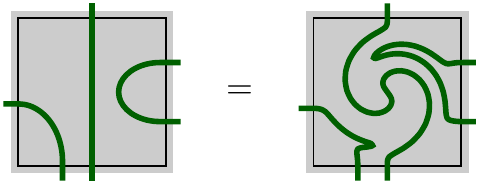}
\end{center}

We will also need to record
various other attributes of these curves,
and to do this we make this notation more elaborate
in the paragraphs {\bf (I)}, {\bf(II)} and {\bf(III)} below.
Each symbol in the word describes an intersection of
the curve with the tile boundary,
and so as we decorate these symbols these decorations will
apply to such intersection points.

\begin{center}
\includegraphics[]{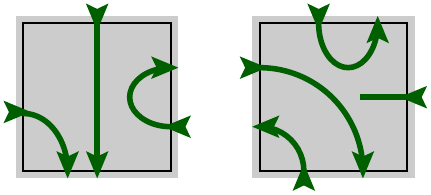}
\end{center}

{\bf (I)} We will record the direction of each piece-of-curve,
this will be either an {\tt in} or {\tt out} decoration for each symbol.
Such decorations need to balance according to the brackets.
The decorated symbols $*_{\mbox{\tt in}}$ and 
$*_{\mbox{\tt out}}$ 
will denote respectively either
the head or the tail of a curve.
The words for the diagrams above now read as
$ \bigl<_{\mbox{\tt in}}\bigl<_{\mbox{\tt out}}\bigr>_{\mbox{\tt in}}
    \bigr>_{\mbox{\tt out}}\bigl<_{\mbox{\tt out}}\bigr>_{\mbox{\tt in}}$
and
$ \bigl<_{\mbox{\tt in}}\bigl>_{\mbox{\tt out}}*_{\mbox{\tt in}}
    \bigr<_{\mbox{\tt out}}
    \bigr<_{\mbox{\tt in}}\bigl>_{\mbox{\tt out}}\bigr>_{\mbox{\tt in}}.
$

{\bf (II)} We will record,
for each intersection with the tile edge, 
a numeral indicating which of the four
sides of the tile the
intersection occurs on.
Numbering these clockwise from the top as $1, 2, 3, 4$ we have for the above curves: 
$\bigl<_1\bigl<_2\bigr>_2\bigr>_3\bigl<_3\bigr>_4$ 
and $\bigl<_1\bigr>_1*_2\bigl<_3\bigl<_3\bigr>_4\bigr>_4.$

{\bf (III)} Finally, we will also decorate these symbols with anyons.
This will be an index to a leaf of a (sum of) fusion tree(s).
This means that anyons only reside on the curve close
to the tile boundary,
and so we cannot have more than two anyons
for each piece-of-curve. 
The number of such pieces is arbitrary, and so this
is no restriction on generality.

\begin{center}
\includegraphics[]{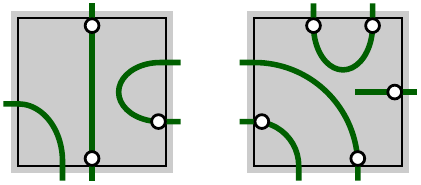}
\end{center}

In joining tiles together to make a tiling we will
require adjacent tiles to agree on their shared boundary.
This will entail sequentially pairing symbols in the
words for adjacent tiles
and requiring that 
the {\tt in} and {\tt out} decorations are matched.
Because the word for a tile proceeds conter-clockwise
around the tile, this pairing will always reverse the
sequential order of the symbols of adjacent tiles.
For example, given the above two tiles we sequentially pair the 
$\bigl<_{\mbox{\tt out},2}\bigr>_{\mbox{\tt in},2}$ 
and $\bigr>_{\mbox{\tt out},4}\bigr>_{\mbox{\tt in},4}$
symbols with opposite order so that
$\bigl<_{\mbox{\tt out},2}\sim\bigr>_{\mbox{\tt in},4}$
and $\bigr>_{\mbox{\tt in},2} \sim \bigr>_{\mbox{\tt out},4}.$ 

Note that in general, this will allow us to store the multiple disjoint curve diagrams that cover the lattice.

For each piece-of-curve, apart from a head or tail, there is an associated 
number we call the \emph{turn number}. This counts the number
of ``right-hand turns'' the piece-of-curve makes as it
traverses the tile, with a ``left-hand turn'' counting as $-1.$
This number will take one of the values $-2, -1, 0, 1, 2:$
\begin{center}
\includegraphics[]{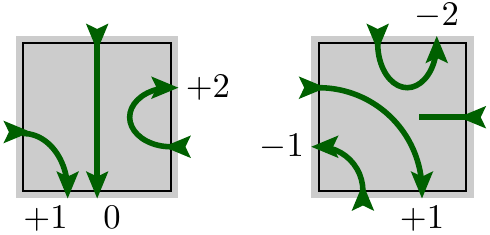}
\end{center}

%
%

\subsection{The paperclip algorithm}

In order to calculate the total charge of each tile, we must transport all anyons within the tile until they are neighbours along their curve. In doing so, it will be convenient to consider transporting these anyons by moving them along tile edges. This is equivalent to a ``refactoring'' of the fusion space, as described in Ref.~\cite{Burton2016}. We can further break down transport along the tile edge into a sequence of transports between neighbouring pieces of curve at that edge.
The origin and destination of such a path
now splits the curve diagram
into three disjoint pieces which we term
\emph{head}, \emph{body} and \emph{tail}, where
the head contains the endpoint of the curve, the tail
contains the beginning of the curve and the body is the remainder.
Here we consider transport forwards along the curve (from tail towards head), but the reverse case is analogous.
\begin{center}
\includegraphics[]{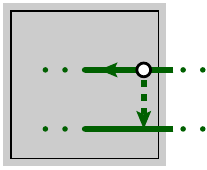}
\end{center}

This arrangement is equivalent (isotopic) to one of four 
``paperclips'', which we distinguish between by counting how
many \emph{right-hand turns} are made along the body of the curve diagram.
We also show an equivalent (isotopic) picture where the
curve diagram has been straightened, and the resulting distortion
in the anyon path:
\begin{center}
\includegraphics[]{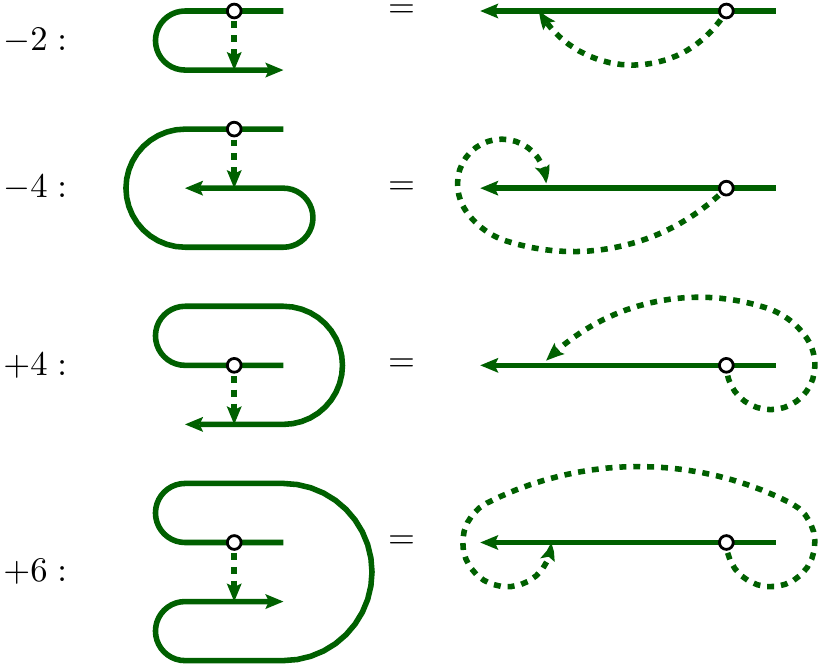}
\end{center}
In order to perform the desired transport, we must braid the anyon under consideration around the other anyons along the curve diagram. Denoting the sequence of anyons along the head, body and tail, $H, B$ and $T,$
respectively, and using
$H^r, B^r$ and $T^r$ to denote the same anyons with the reversed order, the appropriate braiding transformation ($R$-move) can be read off for each
of the four paperclips:
\begin{align*}
-2:&\ R[B] \\
-4:&\ R[H^r]\ R[H]\ R[B] \\
+4:&\ R[B]\ R[T]\ R[T^r] \\
+6:&\ R[H^r]\ R[H]\ R[B]\ R[T]\ R[T^r] \\
\end{align*}
where notation such as $R[B]$ is understood as sequentially clockwise braiding around
each anyon in $B$. 

Once all the anyons within a tile have been brought into a sequential piece of the curve, their total charge can easily be read off by an appropriate $F$-move. This allows us to perform syndrome extraction, by repeating this procedure for each tile of the lattice.

%
%

\subsection{Decoding and error-correction}

After the noise process is applied to the system,
the error correction routine attempts to annihilate all the anyons on the lattice in a way that does not affect the encoded information. It proceeds as a dialogue between the
decoder and the system. The syndrome extraction procedure provides information to the decoder, which suggests a recovery procedure, the results of which are found by a subsequent syndrome extraction step, and so on. This iterative procedure continues until the extracted syndrome is empty (i.e.~i.e. there are no remaining anyons on the lattice), or the decoder declares failure.
Here we show this in a process diagram, with time running up
the page:
\begin{center}
\includegraphics[]{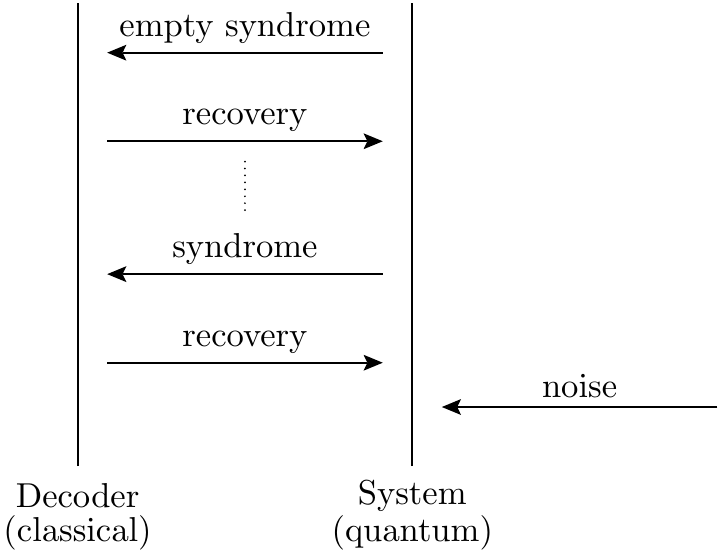}
\end{center}

So far we have discussed the simulation of the (quantum) system
and now we turn to the decoder algorithm. This is similar to other clustering-type decoders used in the literature~\cite{Bravyi2011}, in that it attempts to form and fuse together clusters of anyons on larger and larger length scales, declaring failure if anyons remain after fusion at the largest length scale. Our simulation also declares failure if at any point the curve diagrams describing the state of the system form a homologically non-trivial loop (span the lattice), this being a necessary condition for error-correction failure.

The structure of the error-correction routine is as follows, and will be illustrated by an example below.

\begin{verbatim}
 1:  def error_correct():
 2:      syndrome = get_syndrome()
 3:      
 4:      # build a cluster for each charge
 5:      clusters = [Cluster(charge) for charge in syndrome]
 6:  
 7:      # join any neighbouring clusters
 8:      join(clusters, 1)
 9:      
10:      while clusters:
11:      
12:          # find total charge on each cluster
13:          for cluster in clusters:
14:              fuse_cluster(cluster)
15:      
16:          # discard vacuum clusters
17:          clusters = [cluster for cluster in clusters if non_vacuum(cluster)]
18:      
19:          # grow each cluster by 1 unit
20:          for cluster in clusters:
21:              grow_cluster(cluster, 1)
22:      
23:          # join any intersecting clusters
24:          join(clusters, 0)
25:  
26:      # success !
27:      return True
\end{verbatim}

First, we show the result of the initial call to {\tt get\_syndrome()}, on line 2. This is computed using the syndrome extraction step as described above.
The locations of anyon charges are highlighted in red.
For each of these charges we build a {\tt Cluster}, on line 5.
Each cluster is shown as a gray shaded area.
\begin{center}
\includegraphics[]{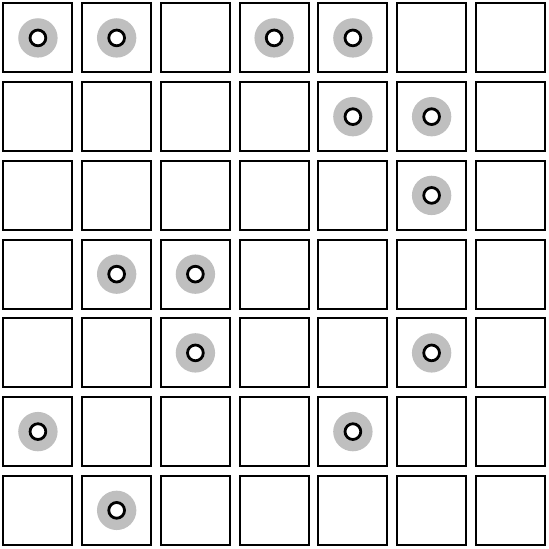}
\end{center}
The next step is the call to {\tt join(clusters, 1)}, on line 8,
which joins clusters that are separated by at most one lattice
spacing. We now have seven clusters:
\begin{center}
\includegraphics[]{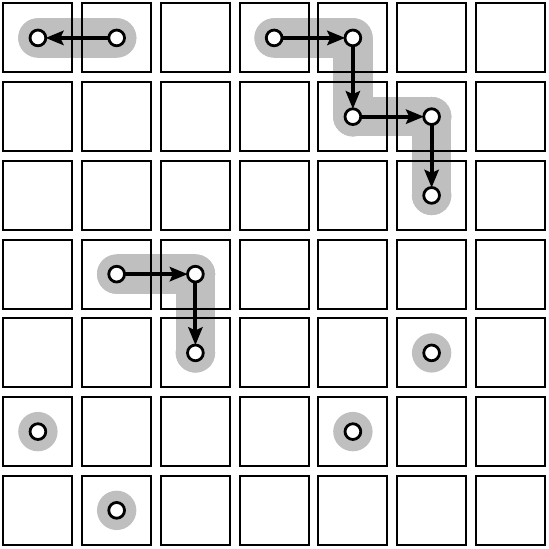}
\end{center}
Each cluster is structured as a rooted tree, as indicated by
the arrows which point in the direction from the leaves to
the root of the tree. 
This tree structure is used in the call to {\tt fuse\_cluster()},
on line 14.
This moves anyons in the tree along the arrows to the root, 
fusing with the charge at the root. This movement of anyons is a physical recovery operation that is easily simulated by extending the curve diagram along the transport path, and moving the anyons along it.
\begin{center}
\includegraphics[]{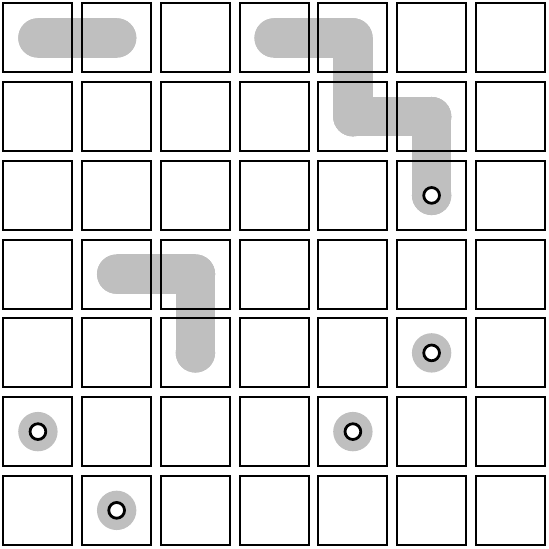}
\end{center}
For each cluster, the charge is then measured using the syndrome extraction procedure, and the resulting charge at the root is taken as the charge of
that cluster. Any cluster with vacuum total charge is then discarded (line 17).
In our example, we find two clusters with vacuum charge and we discard these.
The next step is to grow the remaining clusters by one lattice spacing (line 20-21),
and join (merge) any overlapping clusters (line 24).
\begin{center}
\includegraphics[]{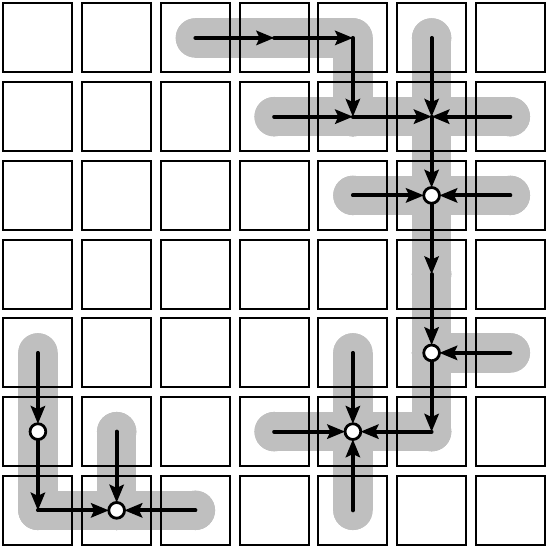}
\end{center}
Note that we can choose the root of each cluster arbitrarily
since we are only interested in the total charge of each cluster, and also that for simplicity we have neglected the boundary of the lattice in
this example.

We repeat these steps of fusing, growing and then joining clusters (lines 10-24.)
If at any point this causes a topologically 
non-trivial operation, the simulation aborts and a failure
to decode is recorded.
Otherwise we eventually run out
of non-vacuum clusters, and the decoder succeeds (line 27).

By repeatedly performing this simulation of noise and error-correction on different sized lattices, and for different noise rates (or memory lifetimes), we are able to estimate the error-correction threshold for this type of system.


\begin{acknowledgments}

We thank A.\ Doherty and R.\ Pfeifer for discussions. 
This work was supported by the ARC via EQuS project number CE11001013, by the US Army Research Office grant numbers W911NF-14-1-0098 and W911NF-14-1-0103, ERC grant QFTCMPS and by the cluster of excellence EXC 201 Quantum Engineering and Space-Time Research. STF also acknowledges support from an ARC Future Fellowship FT130101744. This research was supported in part by Perimeter Institute for Theoretical Physics. Research at Perimeter Institute is supported by the Government of Canada through Industry Canada and by the Province of Ontario through the Ministry of Research and Innovation.

\end{acknowledgments}


\begin{thebibliography}{43}%
\makeatletter
\providecommand \@ifxundefined [1]{%
 \@ifx{#1\undefined}
}%
\providecommand \@ifnum [1]{%
 \ifnum #1\expandafter \@firstoftwo
 \else \expandafter \@secondoftwo
 \fi
}%
\providecommand \@ifx [1]{%
 \ifx #1\expandafter \@firstoftwo
 \else \expandafter \@secondoftwo
 \fi
}%
\providecommand \natexlab [1]{#1}%
\providecommand \enquote  [1]{``#1''}%
\providecommand \bibnamefont  [1]{#1}%
\providecommand \bibfnamefont [1]{#1}%
\providecommand \citenamefont [1]{#1}%
\providecommand \href@noop [0]{\@secondoftwo}%
\providecommand \href [0]{\begingroup \@sanitize@url \@href}%
\providecommand \@href[1]{\@@startlink{#1}\@@href}%
\providecommand \@@href[1]{\endgroup#1\@@endlink}%
\providecommand \@sanitize@url [0]{\catcode `\\12\catcode `\$12\catcode
  `\&12\catcode `\#12\catcode `\^12\catcode `\_12\catcode `\%12\relax}%
\providecommand \@@startlink[1]{}%
\providecommand \@@endlink[0]{}%
\providecommand \url  [0]{\begingroup\@sanitize@url \@url }%
\providecommand \@url [1]{\endgroup\@href {#1}{\urlprefix }}%
\providecommand \urlprefix  [0]{URL }%
\providecommand \Eprint [0]{\href }%
\providecommand \doibase [0]{http://dx.doi.org/}%
\providecommand \selectlanguage [0]{\@gobble}%
\providecommand \bibinfo  [0]{\@secondoftwo}%
\providecommand \bibfield  [0]{\@secondoftwo}%
\providecommand \translation [1]{[#1]}%
\providecommand \BibitemOpen [0]{}%
\providecommand \bibitemStop [0]{}%
\providecommand \bibitemNoStop [0]{.\EOS\space}%
\providecommand \EOS [0]{\spacefactor3000\relax}%
\providecommand \BibitemShut  [1]{\csname bibitem#1\endcsname}%
\let\auto@bib@innerbib\@empty
\bibitem [{\citenamefont {Kitaev}(2003)}]{Kitaev2003}%
  \BibitemOpen
  \bibfield  {author} {\bibinfo {author} {\bibfnamefont {A.~Y.}\ \bibnamefont
  {Kitaev}},\ }\href {\doibase 10.1016/S0003-4916(02)00018-0} {\bibfield
  {journal} {\bibinfo  {journal} {Ann. Phys.}\ }\textbf {\bibinfo {volume}
  {303}},\ \bibinfo {pages} {2} (\bibinfo {year} {2003})},\ \Eprint
  {http://arxiv.org/abs/quant-ph/9707021} {arXiv:quant-ph/9707021} \BibitemShut
  {NoStop}%
\bibitem [{\citenamefont {Dennis}\ \emph {et~al.}(2002)\citenamefont {Dennis},
  \citenamefont {Kitaev}, \citenamefont {Landahl},\ and\ \citenamefont
  {Preskill}}]{Dennis2002}%
  \BibitemOpen
  \bibfield  {author} {\bibinfo {author} {\bibfnamefont {E.}~\bibnamefont
  {Dennis}}, \bibinfo {author} {\bibfnamefont {A.}~\bibnamefont {Kitaev}},
  \bibinfo {author} {\bibfnamefont {A.}~\bibnamefont {Landahl}}, \ and\
  \bibinfo {author} {\bibfnamefont {J.}~\bibnamefont {Preskill}},\ }\href
  {\doibase 10.1063/1.1499754} {\bibfield  {journal} {\bibinfo  {journal} {J.
  Math. Phys.}\ }\textbf {\bibinfo {volume} {43}},\ \bibinfo {pages} {4452}
  (\bibinfo {year} {2002})},\ \Eprint {http://arxiv.org/abs/quant-ph/0110143}
  {arXiv:quant-ph/0110143} \BibitemShut {NoStop}%
\bibitem [{\citenamefont {Nayak}\ \emph {et~al.}(2008)\citenamefont {Nayak},
  \citenamefont {Simon}, \citenamefont {Stern}, \citenamefont {Freedman},\ and\
  \citenamefont {Das~Sarma}}]{Nayak2008}%
  \BibitemOpen
  \bibfield  {author} {\bibinfo {author} {\bibfnamefont {C.}~\bibnamefont
  {Nayak}}, \bibinfo {author} {\bibfnamefont {S.~H.}\ \bibnamefont {Simon}},
  \bibinfo {author} {\bibfnamefont {A.}~\bibnamefont {Stern}}, \bibinfo
  {author} {\bibfnamefont {M.}~\bibnamefont {Freedman}}, \ and\ \bibinfo
  {author} {\bibfnamefont {S.}~\bibnamefont {Das~Sarma}},\ }\href {\doibase
  10.1103/RevModPhys.80.1083} {\bibfield  {journal} {\bibinfo  {journal} {Rev.
  Mod. Phys.}\ }\textbf {\bibinfo {volume} {80}},\ \bibinfo {pages} {1083}
  (\bibinfo {year} {2008})},\ \Eprint {http://arxiv.org/abs/0707.1889}
  {arXiv:0707.1889} \BibitemShut {NoStop}%
\bibitem [{\citenamefont {Bravyi}\ \emph {et~al.}(2010)\citenamefont {Bravyi},
  \citenamefont {Hastings},\ and\ \citenamefont {Michalakis}}]{Bravyi2010}%
  \BibitemOpen
  \bibfield  {author} {\bibinfo {author} {\bibfnamefont {S.}~\bibnamefont
  {Bravyi}}, \bibinfo {author} {\bibfnamefont {M.}~\bibnamefont {Hastings}}, \
  and\ \bibinfo {author} {\bibfnamefont {S.}~\bibnamefont {Michalakis}},\
  }\href {\doibase 10.1063/1.3490195} {\bibfield  {journal} {\bibinfo
  {journal} {J. Math. Phys.}\ }\textbf {\bibinfo {volume} {51}},\ \bibinfo
  {pages} {093512} (\bibinfo {year} {2010})},\ \Eprint
  {http://arxiv.org/abs/1001.0344} {arXiv:1001.0344} \BibitemShut {NoStop}%
\bibitem [{\citenamefont {Bravyi}\ and\ \citenamefont
  {Hastings}(2011)}]{Bravyi2011a}%
  \BibitemOpen
  \bibfield  {author} {\bibinfo {author} {\bibfnamefont {S.}~\bibnamefont
  {Bravyi}}\ and\ \bibinfo {author} {\bibfnamefont {M.~B.}\ \bibnamefont
  {Hastings}},\ }\href {\doibase 10.1007/s00220-011-1346-2} {\bibfield
  {journal} {\bibinfo  {journal} {Comm. Math. Phys.}\ }\textbf {\bibinfo
  {volume} {307}},\ \bibinfo {pages} {609} (\bibinfo {year} {2011})},\ \Eprint
  {http://arxiv.org/abs/arXiv:1001.4363} {arXiv:1001.4363} \BibitemShut
  {NoStop}%
\bibitem [{\citenamefont {Michalakis}\ and\ \citenamefont
  {Zwolak}(2013)}]{Michalakis2013}%
  \BibitemOpen
  \bibfield  {author} {\bibinfo {author} {\bibfnamefont {S.}~\bibnamefont
  {Michalakis}}\ and\ \bibinfo {author} {\bibfnamefont {J.~P.}\ \bibnamefont
  {Zwolak}},\ }\href {\doibase 10.1007/s00220-013-1762-6} {\bibfield  {journal}
  {\bibinfo  {journal} {Comm. Math. Phys.}\ }\textbf {\bibinfo {volume}
  {322}},\ \bibinfo {pages} {277} (\bibinfo {year} {2013})},\ \Eprint
  {http://arxiv.org/abs/1109.1588} {arXiv:1109.1588} \BibitemShut {NoStop}%
\bibitem [{\citenamefont {Wilczek}(1990)}]{Wilczek1990}%
  \BibitemOpen
  \bibfield  {author} {\bibinfo {author} {\bibfnamefont {F.}~\bibnamefont
  {Wilczek}},\ }\href@noop {} {\emph {\bibinfo {title} {Fractional Statistics
  and Anyon Superconductivity}}}\ (\bibinfo  {publisher} {World Scientific},\
  \bibinfo {address} {Singapore},\ \bibinfo {year} {1990})\BibitemShut
  {NoStop}%
\bibitem [{\citenamefont {Freedman}\ \emph {et~al.}(2002)\citenamefont
  {Freedman}, \citenamefont {Larsen},\ and\ \citenamefont
  {Wang}}]{Freedman2002}%
  \BibitemOpen
  \bibfield  {author} {\bibinfo {author} {\bibfnamefont {M.~H.}\ \bibnamefont
  {Freedman}}, \bibinfo {author} {\bibfnamefont {M.}~\bibnamefont {Larsen}}, \
  and\ \bibinfo {author} {\bibfnamefont {Z.}~\bibnamefont {Wang}},\ }\href
  {\doibase 10.1007/s002200200645} {\bibfield  {journal} {\bibinfo  {journal}
  {Comm. Math. Phys.}\ }\textbf {\bibinfo {volume} {227}},\ \bibinfo {pages}
  {605} (\bibinfo {year} {2002})},\ \Eprint
  {http://arxiv.org/abs/quant-ph/0001108} {arXiv:quant-ph/0001108} \BibitemShut
  {NoStop}%
\bibitem [{\citenamefont {Pastawski}\ \emph {et~al.}(2010)\citenamefont
  {Pastawski}, \citenamefont {Kay}, \citenamefont {Schuch},\ and\ \citenamefont
  {Cirac}}]{Pastawski2010}%
  \BibitemOpen
  \bibfield  {author} {\bibinfo {author} {\bibfnamefont {F.}~\bibnamefont
  {Pastawski}}, \bibinfo {author} {\bibfnamefont {A.}~\bibnamefont {Kay}},
  \bibinfo {author} {\bibfnamefont {N.}~\bibnamefont {Schuch}}, \ and\ \bibinfo
  {author} {\bibfnamefont {J.~I.}\ \bibnamefont {Cirac}},\ }\href@noop {}
  {\bibfield  {journal} {\bibinfo  {journal} {Quant. Inf. Comput.}\ }\textbf
  {\bibinfo {volume} {10}},\ \bibinfo {pages} {580} (\bibinfo {year} {2010})},\
  \Eprint {http://arxiv.org/abs/0911.3843} {arXiv:0911.3843} \BibitemShut
  {NoStop}%
\bibitem [{\citenamefont {Duclos-Cianci}\ and\ \citenamefont
  {Poulin}(2010{\natexlab{a}})}]{Duclos-Cianci2010}%
  \BibitemOpen
  \bibfield  {author} {\bibinfo {author} {\bibfnamefont {G.}~\bibnamefont
  {Duclos-Cianci}}\ and\ \bibinfo {author} {\bibfnamefont {D.}~\bibnamefont
  {Poulin}},\ }\href {\doibase 10.1103/PhysRevLett.104.050504} {\bibfield
  {journal} {\bibinfo  {journal} {Phys. Rev. Lett.}\ }\textbf {\bibinfo
  {volume} {104}},\ \bibinfo {pages} {050504} (\bibinfo {year}
  {2010}{\natexlab{a}})},\ \Eprint {http://arxiv.org/abs/0911.0581}
  {arXiv:0911.0581} \BibitemShut {NoStop}%
\bibitem [{\citenamefont {Duclos-Cianci}\ and\ \citenamefont
  {Poulin}(2010{\natexlab{b}})}]{Duclos-Cianci2010a}%
  \BibitemOpen
  \bibfield  {author} {\bibinfo {author} {\bibfnamefont {G.}~\bibnamefont
  {Duclos-Cianci}}\ and\ \bibinfo {author} {\bibfnamefont {D.}~\bibnamefont
  {Poulin}},\ }in\ \href {\doibase 10.1109/CIG.2010.5592866} {\emph {\bibinfo
  {booktitle} {Information Theory Workshop (ITW), 2010 IEEE}}}\ (\bibinfo
  {year} {2010})\ pp.\ \bibinfo {pages} {1--5},\ \Eprint
  {http://arxiv.org/abs/1006.1362} {arXiv:1006.1362} \BibitemShut {NoStop}%
\bibitem [{\citenamefont {Wang}\ \emph
  {et~al.}(2010{\natexlab{a}})\citenamefont {Wang}, \citenamefont {Fowler},
  \citenamefont {Stephens},\ and\ \citenamefont {Hollenberg}}]{Wang2010}%
  \BibitemOpen
  \bibfield  {author} {\bibinfo {author} {\bibfnamefont {D.~S.}\ \bibnamefont
  {Wang}}, \bibinfo {author} {\bibfnamefont {A.~G.}\ \bibnamefont {Fowler}},
  \bibinfo {author} {\bibfnamefont {A.~M.}\ \bibnamefont {Stephens}}, \ and\
  \bibinfo {author} {\bibfnamefont {L.~C.~L.}\ \bibnamefont {Hollenberg}},\
  }\href@noop {} {\bibfield  {journal} {\bibinfo  {journal} {Quant. Inf.
  Comput.}\ }\textbf {\bibinfo {volume} {10}},\ \bibinfo {pages} {456}
  (\bibinfo {year} {2010}{\natexlab{a}})},\ \Eprint
  {http://arxiv.org/abs/0905.0531} {arXiv:0905.0531} \BibitemShut {NoStop}%
\bibitem [{\citenamefont {Wang}\ \emph
  {et~al.}(2010{\natexlab{b}})\citenamefont {Wang}, \citenamefont {Fowler},
  \citenamefont {Hill},\ and\ \citenamefont {Hollenberg}}]{Wang2010a}%
  \BibitemOpen
  \bibfield  {author} {\bibinfo {author} {\bibfnamefont {D.~S.}\ \bibnamefont
  {Wang}}, \bibinfo {author} {\bibfnamefont {A.~G.}\ \bibnamefont {Fowler}},
  \bibinfo {author} {\bibfnamefont {C.~D.}\ \bibnamefont {Hill}}, \ and\
  \bibinfo {author} {\bibfnamefont {L.~C.~L.}\ \bibnamefont {Hollenberg}},\
  }\href@noop {} {\bibfield  {journal} {\bibinfo  {journal} {Quant. Inf.
  Comput.}\ }\textbf {\bibinfo {volume} {10}},\ \bibinfo {pages} {780}
  (\bibinfo {year} {2010}{\natexlab{b}})},\ \Eprint
  {http://arxiv.org/abs/0907.1708} {arXiv:0907.1708} \BibitemShut {NoStop}%
\bibitem [{\citenamefont {Duclos-Cianci}\ and\ \citenamefont
  {Poulin}(2013)}]{Duclos-Cianci2013}%
  \BibitemOpen
  \bibfield  {author} {\bibinfo {author} {\bibfnamefont {G.}~\bibnamefont
  {Duclos-Cianci}}\ and\ \bibinfo {author} {\bibfnamefont {D.}~\bibnamefont
  {Poulin}},\ }\href@noop {} {\bibfield  {journal} {\bibinfo  {journal} {Quant.
  Inf. Comput.}\ } (\bibinfo {year} {2013})},\ \Eprint
  {http://arxiv.org/abs/1304.6100} {arXiv:1304.6100} \BibitemShut {NoStop}%
\bibitem [{\citenamefont {Bravyi}\ and\ \citenamefont
  {Haah}(2013)}]{Bravyi2011}%
  \BibitemOpen
  \bibfield  {author} {\bibinfo {author} {\bibfnamefont {S.}~\bibnamefont
  {Bravyi}}\ and\ \bibinfo {author} {\bibfnamefont {J.}~\bibnamefont {Haah}},\
  }\href {\doibase 10.1103/PhysRevLett.111.200501} {\bibfield  {journal}
  {\bibinfo  {journal} {Phys. Rev. Lett.}\ }\textbf {\bibinfo {volume} {111}},\
  \bibinfo {pages} {200501} (\bibinfo {year} {2013})},\ \Eprint
  {http://arxiv.org/abs/1112.3252} {arXiv:1112.3252} \BibitemShut {NoStop}%
\bibitem [{\citenamefont {Bombin}\ \emph {et~al.}(2012)\citenamefont {Bombin},
  \citenamefont {Duclos-Cianci},\ and\ \citenamefont {Poulin}}]{Bombin2012}%
  \BibitemOpen
  \bibfield  {author} {\bibinfo {author} {\bibfnamefont {H.}~\bibnamefont
  {Bombin}}, \bibinfo {author} {\bibfnamefont {G.}~\bibnamefont
  {Duclos-Cianci}}, \ and\ \bibinfo {author} {\bibfnamefont {D.}~\bibnamefont
  {Poulin}},\ }\href {http://stacks.iop.org/1367-2630/14/i=7/a=073048}
  {\bibfield  {journal} {\bibinfo  {journal} {New J. Phys.}\ }\textbf {\bibinfo
  {volume} {14}},\ \bibinfo {pages} {073048} (\bibinfo {year} {2012})},\
  \Eprint {http://arxiv.org/abs/1103.4606} {arXiv:1103.4606} \BibitemShut
  {NoStop}%
\bibitem [{\citenamefont {Wootton}\ and\ \citenamefont
  {Loss}(2012)}]{Wootton2012}%
  \BibitemOpen
  \bibfield  {author} {\bibinfo {author} {\bibfnamefont {J.~R.}\ \bibnamefont
  {Wootton}}\ and\ \bibinfo {author} {\bibfnamefont {D.}~\bibnamefont {Loss}},\
  }\href {\doibase 10.1103/PhysRevLett.109.160503} {\bibfield  {journal}
  {\bibinfo  {journal} {Phys. Rev. Lett.}\ }\textbf {\bibinfo {volume} {109}},\
  \bibinfo {pages} {160503} (\bibinfo {year} {2012})},\ \Eprint
  {http://arxiv.org/abs/1202.4316} {arXiv:1202.4316} \BibitemShut {NoStop}%
\bibitem [{\citenamefont {Anwar}\ \emph {et~al.}(2014)\citenamefont {Anwar},
  \citenamefont {Brown}, \citenamefont {Campbell},\ and\ \citenamefont
  {Browne}}]{Anwar2014}%
  \BibitemOpen
  \bibfield  {author} {\bibinfo {author} {\bibfnamefont {H.}~\bibnamefont
  {Anwar}}, \bibinfo {author} {\bibfnamefont {B.~J.}\ \bibnamefont {Brown}},
  \bibinfo {author} {\bibfnamefont {E.~T.}\ \bibnamefont {Campbell}}, \ and\
  \bibinfo {author} {\bibfnamefont {D.~E.}\ \bibnamefont {Browne}},\ }\href
  {\doibase 10.1088/1367-2630/16/6/063038} {\bibfield  {journal} {\bibinfo
  {journal} {New J. Phys.}\ }\textbf {\bibinfo {volume} {16}},\ \bibinfo
  {pages} {063038} (\bibinfo {year} {2014})},\ \Eprint
  {http://arxiv.org/abs/1311.4895} {arXiv:1311.4895} \BibitemShut {NoStop}%
\bibitem [{\citenamefont {Watson}\ \emph {et~al.}(2014)\citenamefont {Watson},
  \citenamefont {Anwar},\ and\ \citenamefont {Browne}}]{Watson2014}%
  \BibitemOpen
  \bibfield  {author} {\bibinfo {author} {\bibfnamefont {F.~H.~E.}\
  \bibnamefont {Watson}}, \bibinfo {author} {\bibfnamefont {H.}~\bibnamefont
  {Anwar}}, \ and\ \bibinfo {author} {\bibfnamefont {D.~E.}\ \bibnamefont
  {Browne}},\ }\href {\doibase 10.1103/PhysRevA.92.032309} {\bibfield  {journal}
  {\bibinfo  {journal} {Phys. Rev. A}\ }\textbf {\bibinfo {volume} {92}},\
  \bibinfo {pages} {032309} (\bibinfo {year} {2015})},\ \Eprint
  {http://arxiv.org/abs/1411.3028} {arXiv:1411.3028} \BibitemShut {NoStop}%
\bibitem [{\citenamefont {Hutter}\ \emph {et~al.}(2014)\citenamefont {Hutter},
  \citenamefont {Wootton},\ and\ \citenamefont {Loss}}]{Hutter2014a}%
  \BibitemOpen
  \bibfield  {author} {\bibinfo {author} {\bibfnamefont {A.}~\bibnamefont
  {Hutter}}, \bibinfo {author} {\bibfnamefont {J.~R.}\ \bibnamefont {Wootton}},
  \ and\ \bibinfo {author} {\bibfnamefont {D.}~\bibnamefont {Loss}},\ }\href
  {\doibase 10.1103/PhysRevA.89.022326} {\bibfield  {journal} {\bibinfo
  {journal} {Phys. Rev. A}\ }\textbf {\bibinfo {volume} {89}},\ \bibinfo
  {pages} {022326} (\bibinfo {year} {2014})},\ \Eprint
  {http://arxiv.org/abs/1302.2669} {arXiv:1302.2669} \BibitemShut {NoStop}%
\bibitem [{\citenamefont {Bravyi}\ \emph {et~al.}(2014)\citenamefont {Bravyi},
  \citenamefont {Suchara},\ and\ \citenamefont {Vargo}}]{Bravyi2014}%
  \BibitemOpen
  \bibfield  {author} {\bibinfo {author} {\bibfnamefont {S.}~\bibnamefont
  {Bravyi}}, \bibinfo {author} {\bibfnamefont {M.}~\bibnamefont {Suchara}}, \
  and\ \bibinfo {author} {\bibfnamefont {A.}~\bibnamefont {Vargo}},\ }\href
  {\doibase 10.1103/PhysRevA.90.032326} {\bibfield  {journal} {\bibinfo
  {journal} {Phys. Rev. A}\ }\textbf {\bibinfo {volume} {90}},\ \bibinfo
  {pages} {032326} (\bibinfo {year} {2014})},\ \Eprint
  {http://arxiv.org/abs/1405.4883} {arXiv:1405.4883} \BibitemShut {NoStop}%
\bibitem [{\citenamefont {Wootton}(2015)}]{Wootton2015}%
  \BibitemOpen
  \bibfield  {author} {\bibinfo {author} {\bibfnamefont {J.~R.}\ \bibnamefont
  {Wootton}},\ }\href {\doibase 10.3390/e17041946} {\bibfield  {journal}
  {\bibinfo  {journal} {Entropy}\ }\textbf {\bibinfo {volume} {17}},\ \bibinfo
  {pages} {1946} (\bibinfo {year} {2015})},\ \Eprint
  {http://arxiv.org/abs/1310.2393} {arXiv:1310.2393} \BibitemShut {NoStop}%
\bibitem [{\citenamefont {Fowler}(2015)}]{Fowler2015}%
  \BibitemOpen
  \bibfield  {author} {\bibinfo {author} {\bibfnamefont {A.~G.}\ \bibnamefont
  {Fowler}},\ }\href@noop {} {\bibfield  {journal} {\bibinfo  {journal} {Quant.
  Inf. Comput.}\ }\textbf {\bibinfo {volume} {15}},\ \bibinfo {pages} {0145}
  (\bibinfo {year} {2015})},\ \Eprint {http://arxiv.org/abs/1307.1740}
  {arXiv:1307.1740} \BibitemShut {NoStop}%
\bibitem [{\citenamefont {Andrist}\ \emph {et~al.}(2015)\citenamefont
  {Andrist}, \citenamefont {Wootton},\ and\ \citenamefont
  {Katzgraber}}]{Andrist2015}%
  \BibitemOpen
  \bibfield  {author} {\bibinfo {author} {\bibfnamefont {R.~S.}\ \bibnamefont
  {Andrist}}, \bibinfo {author} {\bibfnamefont {J.~R.}\ \bibnamefont
  {Wootton}}, \ and\ \bibinfo {author} {\bibfnamefont {H.~G.}\ \bibnamefont
  {Katzgraber}},\ }\href {\doibase 10.1103/PhysRevA.91.042331} {\bibfield
  {journal} {\bibinfo  {journal} {Phys. Rev. A}\ }\textbf {\bibinfo {volume}
  {91}},\ \bibinfo {pages} {042331} (\bibinfo {year} {2015})},\ \Eprint
  {http://arxiv.org/abs/1406.5974} {arXiv:1406.5974} \BibitemShut {NoStop}%
\bibitem [{\citenamefont {Brell}\ \emph {et~al.}(2014)\citenamefont {Brell},
  \citenamefont {Burton}, \citenamefont {Dauphinais}, \citenamefont {Flammia},\
  and\ \citenamefont {Poulin}}]{Brell2013}%
  \BibitemOpen
  \bibfield  {author} {\bibinfo {author} {\bibfnamefont {C.~G.}\ \bibnamefont
  {Brell}}, \bibinfo {author} {\bibfnamefont {S.}~\bibnamefont {Burton}},
  \bibinfo {author} {\bibfnamefont {G.}~\bibnamefont {Dauphinais}}, \bibinfo
  {author} {\bibfnamefont {S.~T.}\ \bibnamefont {Flammia}}, \ and\ \bibinfo
  {author} {\bibfnamefont {D.}~\bibnamefont {Poulin}},\ }\href {\doibase
  10.1103/PhysRevX.4.031058} {\bibfield  {journal} {\bibinfo  {journal} {Phys.
  Rev. X}\ }\textbf {\bibinfo {volume} {4}},\ \bibinfo {pages} {031058}
  (\bibinfo {year} {2014})},\ \Eprint {http://arxiv.org/abs/1311.0019}
  {arXiv:1311.0019} \BibitemShut {NoStop}%
\bibitem [{\citenamefont {Wootton}\ \emph {et~al.}(2014)\citenamefont
  {Wootton}, \citenamefont {Burri}, \citenamefont {Iblisdir},\ and\
  \citenamefont {Loss}}]{Wootton2013}%
  \BibitemOpen
  \bibfield  {author} {\bibinfo {author} {\bibfnamefont {J.~R.}\ \bibnamefont
  {Wootton}}, \bibinfo {author} {\bibfnamefont {J.}~\bibnamefont {Burri}},
  \bibinfo {author} {\bibfnamefont {S.}~\bibnamefont {Iblisdir}}, \ and\
  \bibinfo {author} {\bibfnamefont {D.}~\bibnamefont {Loss}},\ }\href {\doibase
  10.1103/PhysRevX.4.011051} {\bibfield  {journal} {\bibinfo  {journal} {Phys.
  Rev. X}\ }\textbf {\bibinfo {volume} {4}},\ \bibinfo {pages} {011051}
  (\bibinfo {year} {2014})},\ \Eprint {http://arxiv.org/abs/1310.3846}
  {arXiv:1310.3846} \BibitemShut {NoStop}%
\bibitem [{\citenamefont {Hutter}\ \emph {et~al.}(2015)\citenamefont {Hutter},
  \citenamefont {Loss},\ and\ \citenamefont {Wootton}}]{Hutter2014}%
  \BibitemOpen
  \bibfield  {author} {\bibinfo {author} {\bibfnamefont {A.}~\bibnamefont
  {Hutter}}, \bibinfo {author} {\bibfnamefont {D.}~\bibnamefont {Loss}}, \ and\
  \bibinfo {author} {\bibfnamefont {J.~R.}\ \bibnamefont {Wootton}},\ }\href
  {http://stacks.iop.org/1367-2630/17/i=3/a=035017} {\bibfield  {journal}
  {\bibinfo  {journal} {New J. Phys.}\ }\textbf {\bibinfo {volume} {17}},\
  \bibinfo {pages} {035017} (\bibinfo {year} {2015})},\ \Eprint
  {http://arxiv.org/abs/1410.4478} {arXiv:1410.4478} \BibitemShut {NoStop}%
\bibitem [{\citenamefont {Wootton}\ and\ \citenamefont
  {Hutter}(2015)}]{Wootton2015b}%
  \BibitemOpen
  \bibfield  {author} {\bibinfo {author} {\bibfnamefont {J.~R.}\ \bibnamefont
  {Wootton}}\ and\ \bibinfo {author} {\bibfnamefont {A.}~\bibnamefont
  {Hutter}},\ }\href {\doibase 	10.1103/PhysRevA.93.022318} {\bibfield
  {journal} {\bibinfo  {journal} {Phys. Rev. A}\ }\textbf {\bibinfo {volume}
  {93}},\ \bibinfo {pages} {042318} (\bibinfo {year} {2016})},\ \Eprint
  {http://arxiv.org/abs/1506.00524} {arXiv:1506.00524} \BibitemShut {NoStop}%
\bibitem [{\citenamefont {Hutter}\ and\ \citenamefont
  {Wootton}(2015)}]{Hutter2015continuous}%
  \BibitemOpen
  \bibfield  {author} {\bibinfo {author} {\bibfnamefont {A.}~\bibnamefont
  {Hutter}}\ and\ \bibinfo {author} {\bibfnamefont {J.~R.}\ \bibnamefont
  {Wootton}},\ }\href {\doibase 	10.1103/PhysRevA.93.042327} {\bibfield
  {journal} {\bibinfo  {journal} {Phys. Rev. A}\ }\textbf {\bibinfo {volume}
  {93}},\ \bibinfo {pages} {042327} (\bibinfo {year} {2016})},\ \Eprint
  {http://arxiv.org/abs/1508.04033} {arXiv:1508.04033} \BibitemShut {NoStop}%
\bibitem [{\citenamefont {Slingerland}\ and\ \citenamefont
  {Bais}(2001)}]{Slingerland2001}%
  \BibitemOpen
  \bibfield  {author} {\bibinfo {author} {\bibfnamefont {J.~K.}\ \bibnamefont
  {Slingerland}}\ and\ \bibinfo {author} {\bibfnamefont {F.~A.}\ \bibnamefont
  {Bais}},\ }\href {\doibase 10.1016/S0550-3213(01)00308-X} {\bibfield
  {journal} {\bibinfo  {journal} {Nucl. Phys. B}\ }\textbf {\bibinfo {volume}
  {612}},\ \bibinfo {pages} {229} (\bibinfo {year} {2001})},\ \Eprint
  {http://arxiv.org/abs/cond-mat/0104035} {arXiv:cond-mat/0104035} \BibitemShut
  {NoStop}%
\bibitem [{\citenamefont {Levin}\ and\ \citenamefont {Wen}(2005)}]{Levin2005}%
  \BibitemOpen
  \bibfield  {author} {\bibinfo {author} {\bibfnamefont {M.~A.}\ \bibnamefont
  {Levin}}\ and\ \bibinfo {author} {\bibfnamefont {X.-G.}\ \bibnamefont
  {Wen}},\ }\href {\doibase 10.1103/PhysRevB.71.045110} {\bibfield  {journal}
  {\bibinfo  {journal} {Phys. Rev. B}\ }\textbf {\bibinfo {volume} {71}},\
  \bibinfo {pages} {045110} (\bibinfo {year} {2005})},\ \Eprint
  {http://arxiv.org/abs/cond-mat/0404617} {arXiv:cond-mat/0404617} \BibitemShut
  {NoStop}%
\bibitem [{\citenamefont {Bonesteel}\ and\ \citenamefont
  {DiVincenzo}(2012)}]{Bonesteel2012}%
  \BibitemOpen
  \bibfield  {author} {\bibinfo {author} {\bibfnamefont {N.~E.}\ \bibnamefont
  {Bonesteel}}\ and\ \bibinfo {author} {\bibfnamefont {D.~P.}\ \bibnamefont
  {DiVincenzo}},\ }\href {\doibase 10.1103/PhysRevB.86.165113} {\bibfield
  {journal} {\bibinfo  {journal} {Phys. Rev. B}\ }\textbf {\bibinfo {volume}
  {86}},\ \bibinfo {pages} {165113} (\bibinfo {year} {2012})},\ \Eprint
  {http://arxiv.org/abs/1206.6048} {arXiv:1206.6048} \BibitemShut {NoStop}%
\bibitem [{\citenamefont {Kapit}\ and\ \citenamefont
  {Simon}(2013)}]{Kapit2013}%
  \BibitemOpen
  \bibfield  {author} {\bibinfo {author} {\bibfnamefont {E.}~\bibnamefont
  {Kapit}}\ and\ \bibinfo {author} {\bibfnamefont {S.}~\bibnamefont {Simon}},\
  }\href {\doibase 10.1103/PhysRevB.88.184409} {\bibfield  {journal} {\bibinfo
  {journal} {Phys. Rev. B}\ }\textbf {\bibinfo {volume} {88}},\ \bibinfo
  {pages} {184409} (\bibinfo {year} {2013})},\ \Eprint
  {http://arxiv.org/abs/1307.3485} {arXiv:1307.3485} \BibitemShut {NoStop}%
\bibitem [{\citenamefont {Palumbo}\ and\ \citenamefont
  {Pachos}(2014)}]{Palumbo2014}%
  \BibitemOpen
  \bibfield  {author} {\bibinfo {author} {\bibfnamefont {G.}~\bibnamefont
  {Palumbo}}\ and\ \bibinfo {author} {\bibfnamefont {J.~K.}\ \bibnamefont
  {Pachos}},\ }\href {\doibase 10.1103/PhysRevD.90.027703} {\bibfield
  {journal} {\bibinfo  {journal} {Phys. Rev. D}\ }\textbf {\bibinfo {volume}
  {90}},\ \bibinfo {pages} {027703} (\bibinfo {year} {2014})},\ \Eprint
  {http://arxiv.org/abs/1311.2871} {arXiv:1311.2871} \BibitemShut {NoStop}%
\bibitem [{\citenamefont {Mong}\ \emph {et~al.}(2014)\citenamefont {Mong},
  \citenamefont {Clarke}, \citenamefont {Alicea}, \citenamefont {Lindner},
  \citenamefont {Fendley}, \citenamefont {Nayak}, \citenamefont {Oreg},
  \citenamefont {Stern}, \citenamefont {Berg}, \citenamefont {Shtengel},\ and\
  \citenamefont {Fisher}}]{Mong2014}%
  \BibitemOpen
  \bibfield  {author} {\bibinfo {author} {\bibfnamefont {R.~S.~K.}\
  \bibnamefont {Mong}}, \bibinfo {author} {\bibfnamefont {D.~J.}\ \bibnamefont
  {Clarke}}, \bibinfo {author} {\bibfnamefont {J.}~\bibnamefont {Alicea}},
  \bibinfo {author} {\bibfnamefont {N.~H.}\ \bibnamefont {Lindner}}, \bibinfo
  {author} {\bibfnamefont {P.}~\bibnamefont {Fendley}}, \bibinfo {author}
  {\bibfnamefont {C.}~\bibnamefont {Nayak}}, \bibinfo {author} {\bibfnamefont
  {Y.}~\bibnamefont {Oreg}}, \bibinfo {author} {\bibfnamefont {A.}~\bibnamefont
  {Stern}}, \bibinfo {author} {\bibfnamefont {E.}~\bibnamefont {Berg}},
  \bibinfo {author} {\bibfnamefont {K.}~\bibnamefont {Shtengel}}, \ and\
  \bibinfo {author} {\bibfnamefont {M.~P.~A.}\ \bibnamefont {Fisher}},\ }\href
  {\doibase 10.1103/PhysRevX.4.011036} {\bibfield  {journal} {\bibinfo
  {journal} {Phys. Rev. X}\ }\textbf {\bibinfo {volume} {4}},\ \bibinfo {pages}
  {011036} (\bibinfo {year} {2014})},\ \Eprint {http://arxiv.org/abs/1307.4403}
  {arXiv:1307.4403} \BibitemShut {NoStop}%
\bibitem [{\citenamefont {Wang}(2010)}]{Wang2010b}%
  \BibitemOpen
  \bibfield  {author} {\bibinfo {author} {\bibfnamefont {Z.}~\bibnamefont
  {Wang}},\ }\href@noop {} {\emph {\bibinfo {title} {Topological quantum
  computation}}}\ (\bibinfo  {publisher} {American Mathematical Society},\
  \bibinfo {year} {2010})\BibitemShut {NoStop}%
\bibitem [{\citenamefont {Burton}(2016)}]{Burton2016}%
  \BibitemOpen
  \bibfield  {author} {\bibinfo {author} {\bibfnamefont {S.}~\bibnamefont
  {Burton}},\ }\href {http://arxiv.org/abs/1610.05384} {\  (\bibinfo {year}
  {2016})},\ \Eprint {http://arxiv.org/abs/arXiv:1610.05384} {arXiv:1610.05384}
  \BibitemShut {NoStop}%
\bibitem [{\citenamefont {Hastie}\ \emph {et~al.}(2009)\citenamefont {Hastie},
  \citenamefont {Tibshirani},\ and\ \citenamefont {Friedman}}]{Hastie2009}%
  \BibitemOpen
  \bibfield  {author} {\bibinfo {author} {\bibfnamefont {T.}~\bibnamefont
  {Hastie}}, \bibinfo {author} {\bibfnamefont {R.}~\bibnamefont {Tibshirani}},
  \ and\ \bibinfo {author} {\bibfnamefont {J.}~\bibnamefont {Friedman}},\
  }\href@noop {} {\emph {\bibinfo {title} {The elements of statistical
  learning}}}\ (\bibinfo  {publisher} {Springer},\ \bibinfo {year}
  {2009})\BibitemShut {NoStop}%
\bibitem [{\citenamefont {Bazant}(2000)}]{Bazant2000}%
  \BibitemOpen
  \bibfield  {author} {\bibinfo {author} {\bibfnamefont {M.~Z.}\ \bibnamefont
  {Bazant}},\ }\href {\doibase 10.1103/PhysRevE.62.1660} {\bibfield  {journal}
  {\bibinfo  {journal} {Phys. Rev. E}\ }\textbf {\bibinfo {volume} {62}},\
  \bibinfo {pages} {1660} (\bibinfo {year} {2000})},\ \Eprint
  {http://arxiv.org/abs/cond-mat/9905191} {arXiv:cond-mat/9905191} \BibitemShut
  {NoStop}%
\bibitem [{\citenamefont {Grimmett}(1989)}]{Grimmett1989}%
  \BibitemOpen
  \bibfield  {author} {\bibinfo {author} {\bibfnamefont {G.}~\bibnamefont
  {Grimmett}},\ }\href@noop {} {\emph {\bibinfo {title} {Percolation}}},\
  \bibinfo {edition} {2nd}\ ed.\ (\bibinfo  {publisher} {Springer},\ \bibinfo
  {year} {1989})\BibitemShut {NoStop}%
\bibitem [{\citenamefont {Pfeifer}\ \emph {et~al.}(2012)\citenamefont
  {Pfeifer}, \citenamefont {Buerschaper}, \citenamefont {Trebst}, \citenamefont
  {Ludwig}, \citenamefont {Troyer},\ and\ \citenamefont {Vidal}}]{Pfeifer2012}%
  \BibitemOpen
  \bibfield  {author} {\bibinfo {author} {\bibfnamefont {R.~N.~C.}\
  \bibnamefont {Pfeifer}}, \bibinfo {author} {\bibfnamefont {O.}~\bibnamefont
  {Buerschaper}}, \bibinfo {author} {\bibfnamefont {S.}~\bibnamefont {Trebst}},
  \bibinfo {author} {\bibfnamefont {A.~W.~W.}\ \bibnamefont {Ludwig}}, \bibinfo
  {author} {\bibfnamefont {M.}~\bibnamefont {Troyer}}, \ and\ \bibinfo {author}
  {\bibfnamefont {G.}~\bibnamefont {Vidal}},\ }\href {\doibase
  10.1103/PhysRevB.86.155111} {\bibfield  {journal} {\bibinfo  {journal} {Phys.
  Rev. B}\ }\textbf {\bibinfo {volume} {86}},\ \bibinfo {pages} {155111}
  (\bibinfo {year} {2012})},\ \Eprint {http://arxiv.org/abs/1005.5486}
  {arXiv:1005.5486} \BibitemShut {NoStop}%
\bibitem [{\citenamefont {Hastings}\ \emph {et~al.}(2014)\citenamefont
  {Hastings}, \citenamefont {Watson},\ and\ \citenamefont
  {Melko}}]{Hastings2014}%
  \BibitemOpen
  \bibfield  {author} {\bibinfo {author} {\bibfnamefont {M.~B.}\ \bibnamefont
  {Hastings}}, \bibinfo {author} {\bibfnamefont {G.~H.}\ \bibnamefont
  {Watson}}, \ and\ \bibinfo {author} {\bibfnamefont {R.~G.}\ \bibnamefont
  {Melko}},\ }\href {\doibase 10.1103/PhysRevLett.112.070501} {\bibfield
  {journal} {\bibinfo  {journal} {Phys. Rev. Lett.}\ }\textbf {\bibinfo
  {volume} {112}},\ \bibinfo {pages} {070501} (\bibinfo {year} {2014})},\
  \Eprint {http://arxiv.org/abs/1309.2680} {arXiv:1309.2680} \BibitemShut
  {NoStop}%
\bibitem [{\citenamefont {Abramsky}(2007)}]{Abramsky2007}%
  \BibitemOpen
  \bibfield  {author} {\bibinfo {author} {\bibfnamefont {S.}~\bibnamefont
  {Abramsky}},\ }in\ \href@noop {} {\emph {\bibinfo {booktitle} {Mathematics of
  Quantum Computing and Technology}}},\ \bibinfo {editor} {edited by\ \bibinfo
  {editor} {\bibfnamefont {L.~K.}\ \bibnamefont {G.~Chen}}\ and\ \bibinfo
  {editor} {\bibfnamefont {S.}~\bibnamefont {Lomonaco}}}\ (\bibinfo
  {publisher} {Taylor \& Francis},\ \bibinfo {year} {2007})\ pp.\ \bibinfo
  {pages} {415--458}\BibitemShut {NoStop}%
\end{thebibliography}


%


\end{document}